\documentclass[aps,prb,reprint,superscriptaddress,amsmath,amssymb]{revtex4-2}
\usepackage{graphicx}
\usepackage{dcolumn}
\usepackage{bm}
\usepackage{color}

\begin{document}


\title{Helicity Selection of the Cycloidal Order in Noncentrosymmetric EuIrGe$_3$}

\author{Kenshin Kurauchi}
\affiliation{Department of Quantum Matter, ADSE, Hiroshima University, Higashi-Hiroshima 739-8530, Japan}
\author{Takeshi Matsumura}
\email[]{tmatsu@hiroshima-u.ac.jp}
\affiliation{Department of Quantum Matter, ADSE, Hiroshima University, Higashi-Hiroshima 739-8530, Japan}
\author{Mitsuru Tsukagoshi}
\affiliation{Department of Quantum Matter, ADSE, Hiroshima University, Higashi-Hiroshima 739-8530, Japan}
\author{Nonoka Higa}
\affiliation{Department of Quantum Matter, ADSE, Hiroshima University, Higashi-Hiroshima 739-8530, Japan}
\author{Masashi Kakihana}
\affiliation{Faculty of Science, University of the Ryukyus, Nishihara, Okinawa 903-0213, Japan}
\author{Masato Hedo}
\affiliation{Faculty of Science, University of the Ryukyus, Nishihara, Okinawa 903-0213, Japan}
\author{Takao Nakama}
\affiliation{Faculty of Science, University of the Ryukyus, Nishihara, Okinawa 903-0213, Japan}
\author{Yoshichika \={O}nuki}
\affiliation{Faculty of Science, University of the Ryukyus, Nishihara, Okinawa 903-0213, Japan}
\affiliation{RIKEN Center for Emergent Matter Science, Wako, Saitama 351-0198, Japan}



\begin{abstract}
The magnetic helicities of the cycloidal ordering in EuIrGe$_3$, with a noncentrosymmetric tetragonal structure, have been studied by circularly polarized resonant X-ray diffraction. It is shown that the helicity of each cycloidal domain is uniquely determined and satisfies the symmetry relations of the $C_{4v}$ point group of the crystal structure. 
The result shows that the cycloidal helicity is determined by the Dzyaloshinskii-Moriya type antisymmetric exchange interaction. 
The domain selection and the phase transition by the external magnetic field along [100] and [110] have also been studied. 
It is shown that the cycloidal plane prefers to be perpendicular to the field and the transverse conical state is realized. 
\end{abstract}

\maketitle

Nontrivial magnetic structures emergent in magnetic fields in noncentrosymmetric crystals have been attracting wide interest~\cite{Bogdanov94,Togawa16,Tokura21}. 
They are realized as a result of competing interactions among symmetric and antisymmetric exchange interactions and the Zeeman energy by an external magnetic field ($\bm{H}$), where the Dzyaloshinskii-Moriya (DM) type antisymmetric interaction between spins $\bm{S}_i$ and $\bm{S}_j$ in the form of $\bm{D}_{ij}\cdot (\bm{S}_i \times \bm{S}_j)$ plays a fundamental role~\cite{Dzyaloshinsky58,Moriya60}. 
$\bm{D}_{ij}$ is a constant vector. 
A typical case is the lattice formation of topologically protected spin swirling objects, e.g., magnetic skyrmion lattices in cubic chiral magnets such as MnSi and EuPtSi~\cite{Muhlbauer09,Kakihana18,Kakihana19,Kaneko19,Tabata19,Homma19,Sakakibara19,Takeuchi19,Takeuchi20,Sakakibara21}, 
and chiral soliton lattices in hexagonal chiral magnets such as CrNb$_3$S$_6$ and Yb(Ni,Cu)$_3$Al$_9$~\cite{Togawa12,Honda20,Ohara14,Matsumura17,Ninomiya18,Aoki18,Tsukagoshi23}. 
The spin helicity, i.e., the sense of rotation of the helical ordering in these chiral structures, which is uniquely determined by the $\bm{D}$ vector, is the source of such topological stability. In noncentrosymmetric polar magnets with mirror reflection planes, a N\'eel-type skyrmion lattice can be formed, which is based on a cycloidal magnetic structure~\cite{Kezsmarki15,Kurumaji17}. 
The spiral ordering and the emergent structures in magnetic fields are strongly associated with the symmetry of the crystal. 

EuIrGe$_3$, with the BaNiSn$_3$-type body-centered tetragonal structure (space group $I4mm$), is a noncentrosymmetric polar magnet with a four-fold $c$ axis and mirror reflection planes including the $c$ axis. The point group $C_{4v}$ allows a DM-type antisymmetric exchange interaction. 
In spite of weak magnetic anisotropy of Eu$^{2+}$ ($L=0$, $S=7/2$), EuIrGe$_3$ exhibits successive phase transitions~\cite{Maurya16,Kakihana17,Utsumi18}. 
In a previous paper, we have investigated the magnetic structure of EuIrGe$_3$ in the three successive phases by neutron diffraction and resonant X-ray diffraction (RXD)~\cite{Matsumura22}. 
It was concluded that a longitudinal sinusoidal structure develops below $T_{\text{N}}=12.2$ K with a propagation vector $\bm{q}=(0, 0, 0.792)$ (phase I), which changes to a cycloid with $\bm{q}=(\delta', 0, 0.8)$ below $T_{\text{N}}^{\;\prime}=7.0$ K ($\delta'=0.017$, phase II). 
Below $T_{\text{N}}^{\;*}=5.0$ K, the cycloidal plane rotates by $45^{\circ}$ and the $\bm{q}$ vector changes to $(\delta, \delta, 0.8)$ ($\delta=0.012$, phase III). 
Here, the magnetic moment of Eu on the $j$th lattice point at $\bm{r}_j$ is generally expressed as
\begin{equation}
  \bm{m}(\bm{r}_j) = \bm{m}_{\bm{q}} e^{i \bm{q}\cdot \bm{r}_j}+\text{c.c.}
\end{equation}
where $\bm{m}_{\bm{q}}$ represents the Fourier component of the magnetic structure. 
The magnetic structure of EuIrGe$_3$ is expressed by $\bm{m}_{\bm{q}}=(0, 0, 1)$ in phase I, 
$\bm{m}_{\bm{q}}=(1, 0, \pm i)$ in phase II, and 
$\bm{m}_{\bm{q}}=(1/\sqrt{2}, 1/\sqrt{2}, \pm i)$ in phase III, where the sign of $\pm i$ represents the sense of rotation. 
Since the $\bm{D}$ vector of the antisymmetric exchange interaction is perpendicular to the mirror plane, on which the $\bm{q}$ vector of EuIrGe$_3$ lies, a unique sense of cycloidal rotation should be selected~\cite{Yambe22}. 
However,  the sign of $\pm i$  remains indeterminate in phases II and III. 

In this paper, we report the helicity selection of the cycloid in EuIrGe$_3$, which has been clarified by RXD using circularly polarized beam. 
We also study the cycloidal domain selection in magnetic fields applied along [100] and [110] directions and show that the cycloidal plane prefers to be perpendicular to the magnetic field, which is equivalent to the $\bm{q}\perp\bm{H}$ condition, and the transverse conical state is realized. 


A single crystal of EuIrGe$_3$ was prepared by the In-flux method as described in Ref. \onlinecite{Kakihana17}. 
The electrical resistivity, specific heat, magnetic susceptibility, and magnetization have already been reported, except for the physical properties in magnetic fields along the [110] axis~\cite{Maurya16,Kakihana17}. 
RXD experiment was performed at BL-3A of the Photon Factory, KEK, Japan, using the same sample as in Ref. \onlinecite{Matsumura22}. 
The scattering geometry is shown in the supplemental material (SM)~\cite{SM}. 
Magnetic field was applied by using a vertical field 8 T superconducting cryomagnet. The sample was mounted on a rotating stage so that we could perform measurements for $H\parallel [100]$ and $H \parallel [110]$ by rotating the sample about the $c$ axis. 
We used X-ray energies around the $L_2$ absorption edge of Eu. 
X-ray energy dependence of a resonant magnetic Bragg peak in phase I at 7.5 K is shown in the SM~\cite{SM}. 
Magnetization measurements for $H \parallel [110]$ were performed by using a Quantum Design SQUID magnetometer.


\begin{figure}[t]
  \begin{center}
  \includegraphics[width=8.5cm]{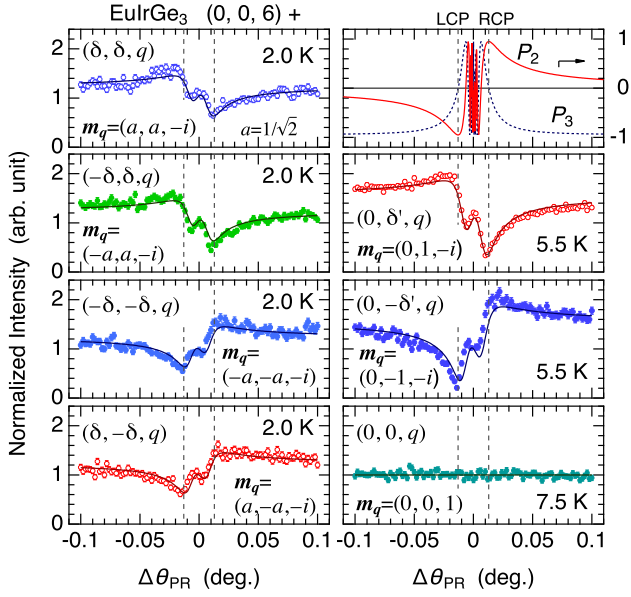}
  \end{center}
  \caption{
(Color online) 
$\Delta \theta_{\text{PR}}(=\theta_{\text{PR}}-\theta_{\text{B}})$ dependences of the intensity of the resonant Bragg peaks at 
$(\pm \delta, \pm \delta , q)$ and $(\pm \delta, \mp \delta, q)$  at 2.0 K in phase III ($\delta = 0.012$, $q=0.8$), 
$(0, \pm \delta ', q)$ at 5.5 K in phase II ($\delta' = 0.017$, $q=0.796$), and $(0, 0, q)$ at 7.5 K in phase I ($q=0.792$), around the (0, 0, 6) fundamental lattice peak. 
The magnetic field is zero. The [100] axis points upward. The background has been subtracted. 
The data are normalized at $\Delta \theta_{\text{PR}}=0$. 
The solid lines are calculations assuming a cycloidal magnetic structure with a single-helicity Fourier component $\bm{m}_{\bm{q}}$ shown in each panel. The X-ray energy is 7.614 keV at resonance. 
The vertical dashed lines represent the positions of the circularly polarized state. 
The right top panel shows the $\Delta \theta_{\text{PR}}$ dependence of the Stokes parameters $P_2$ and $P_3$. 
}
\label{fig:PRscans}
\end{figure}
To determine the cycloidal helicity in phases II and III at zero field, we measured the circular-polarization dependence of the intensities of the resonant magnetic Bragg peaks by using a diamond phase-retarder, which is inserted in the incident beam path. 
By rotating the angle ($\theta_{\text{PR}}$) of the diamond phase plate around the 111 Bragg-angle ($\theta_{\text{B}}$), a phase difference occurs between the $\sigma$ and $\pi$ components. 
The right handed circular polarization (RCP) and left handed circular polarization (LCP) is obtained when the phase difference is equal to $\pm\pi/2$. 
The right top panel of Fig.~\ref{fig:PRscans} shows the offset-angle ($\Delta \theta_{\text{PR}}=\theta_{\text{PR}}-\theta_{\text{B}}$) dependence of the polarization state of the incident beam using the Stokes parameters $P_2$ and $P_3$, where $P_2$ and $P_3$ represent the degree of circular polarization ($+1$ for RCP and $-1$ for LCP)  and linear polarization ($+1$ for $\sigma$ and $-1$ for $\pi$), respectively. 

The data at 2.0 K in Fig.~\ref{fig:PRscans} are the results for the four resonant Bragg peaks in phase III corresponding to the four cycloidal magnetic domains, which clearly demonstrate the asymmetry with respect to the RCP and LCP incident beams. 
The solid lines are the calculations for the $E1$ resonant scattering from cycloidal magnetic structure with a single helicity, i.e., $F\propto (\bm{\varepsilon}' \times \bm{\varepsilon}) \cdot \bm{m}$~\cite{Hannon88,Lovesey05,Nagao05,Nagao06,SM}, where the Fourier component $\bm{m}_{\bm{q}}$ is shown in each panel. 
Since the data are well explained by the calculation, the analysis clearly show that the cycloidal structure in phase III has a single helicity. 
Furthermore, it is noted that the magnetic structures of the four domains perfectly reflect the $C_{4v}$ symmetry of the crystal structure. 
All the four $\bm{m}_{\bm{q}}$ vectors are related by the $90^{\circ}$ rotations and also by the mirror reflections of $C_{4v}$. 
This result clearly show that the helicity of the cycloid is uniquely selected by the DM-type antisymmetric interaction. 

The same analysis can also be applied to the data at 5.5 K in phase II. 
It is noted, however, the $\Delta \theta_{\text{PR}}$ dependence of the two peaks at $(\pm \delta', 0, q)$ becomes flat because of the geometrical reason, where the cycloidal plane lies in the horizontal scattering plane, resulting in $F_{\pi\pi'}=0$ and $|F_{\pi\sigma'}|=|F_{\sigma\pi'}|$. 
The data at 7.5 K in phase I with the longitudinal sinusoidal structure is also flat for the same reason.


\begin{figure}[t]
  \begin{center}
  \includegraphics[width=7.6 cm]{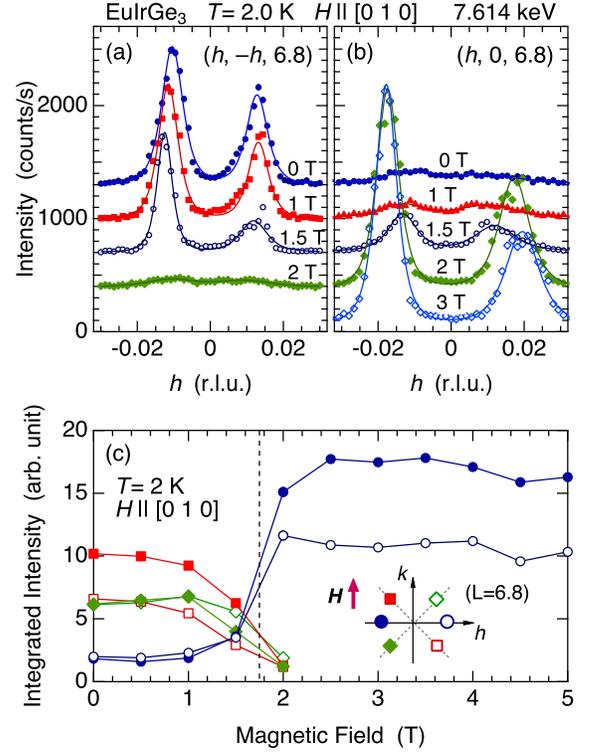}
  \end{center}
  \caption{
(Color online) Magnetic field dependence of the reciprocal space scan along (a) $(h, -h, 6.8)$  and (b) $(h, 0, 6.8)$ at 2.0 K in phase III. 
The magnetic field is applied along [010]. The incident X-ray energy is 7.614 keV at resonance. 
(c) Magnetic field dependence of the integrated intensities of the Bragg peaks at $(\pm \delta, \pm \delta, 6.8)$, $(\pm \delta, \mp \delta, 6.8)$, and $(\pm \delta', 0, 6.8)$. The inset schematically shows the positions of the Bragg peaks.
}
\label{fig:H010domain}
\end{figure}

\begin{figure}[t]
\begin{center}
\includegraphics[width=7.5cm]{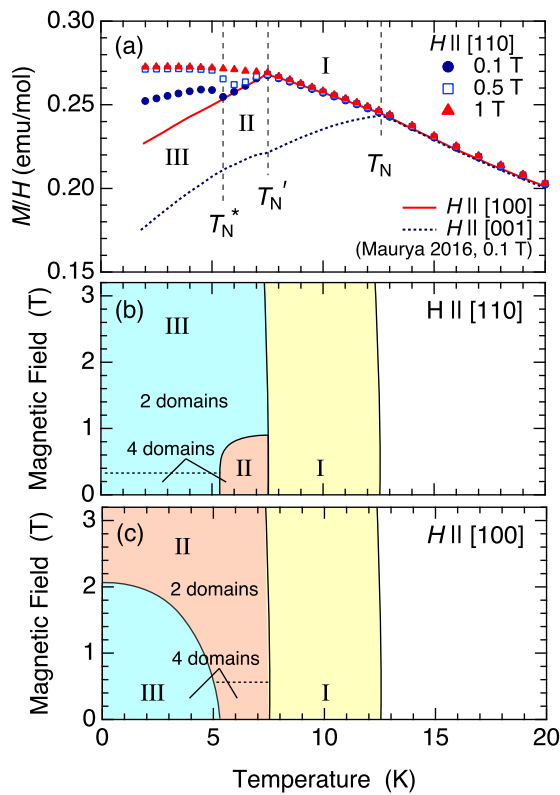}
\end{center}
\caption{
(Color online) Temperature dependences of magnetic susceptibilities $M/H$ of EuIrGe$_3$ at field 1 kG. The magnetic fields along [100] and [001] are reproduced from the literature. 
(a) Temperature dependences of the magnetic susceptibility $M/H$ for $H \parallel [110]$ measured at 0.1 T, 0.5 T, and 1.0 T. 
$M/H$ data for $H \parallel [100]$ and $[001]$ reproduced from the literature are also shown. 
(b) and (c) Magnetic phase diagram for $H \parallel [1 1 0]$ and $[1 0 0]$, respectively, at low magnetic fields below 3 T. 
}
\label{fig:PhaseDiagram}
\end{figure}

When a magnetic field is applied parallel to the $ab$-plane, the magnetic domains are selected so that the cycloidal plane become perpendicular to the applied field.
Figures \ref{fig:H010domain}(a) and \ref{fig:H010domain}(b) show the results of the reciprocal space scans of $(h, -h, 6.8)$ and $(h, 0, 6.8)$, respectively, at several magnetic fields applied along $[0 1 0]$ at 2.0 K in the field increasing process. 
Figure \ref{fig:H010domain}(c) shows the magnetic-field dependences of the integrated intensities of the Bragg peaks, including those obtained from the $(h, h, 6.8)$ scans. 
All the four intensities of the Bragg peaks in phase III decrease with increasing the field and disappear at $\sim 2$ T. 
On the other hand, the two peaks at $(\pm \delta', 0, 6.8)$, corresponding to the cycloid in phase II, appear at high fields above $\sim 2$ T. 
The other two peaks at $(0, \pm \delta', 6.8)$, which weakly exist at 0 T, also vanishes by applying a field. 
These results show that a magnetic phase transition takes place from phase III with four domains to phase II with two domains. 
The two peaks in phase II are at $(\pm \delta', 0, 6.8)$ for $H \parallel [010]$. This means that the cycloidal plane is in the $[1 0 0]$--$[0 0 1]$ plane and the uniform magnetization is induced along the [010] axis. The magnetic phase diagram will be shown later in Fig.~\ref{fig:PhaseDiagram}. 

\begin{figure}[t]
  \begin{center}
  \includegraphics[width=7.6cm]{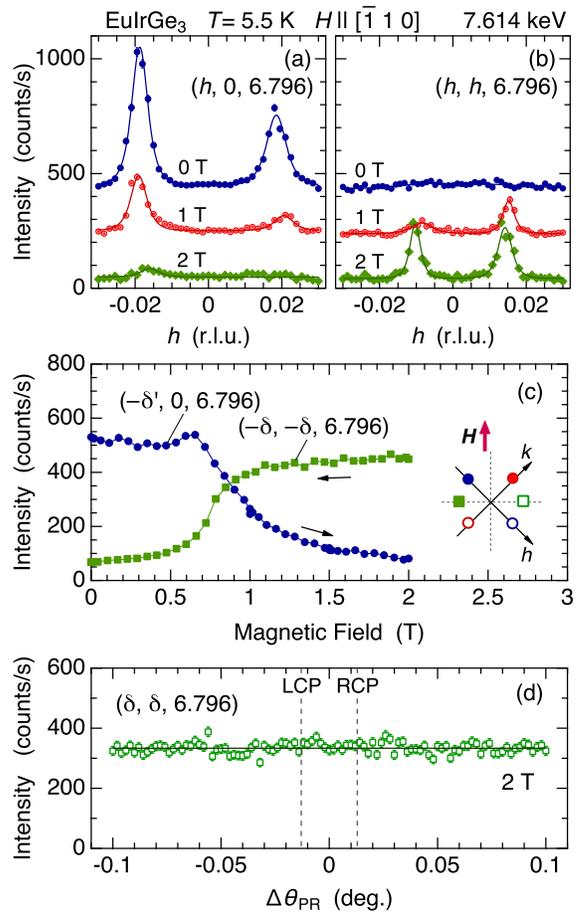}
  \end{center}
  \caption{
(Color online) Magnetic field dependence of the reciprocal space scans along (a) $(h, 0, 6+q)$ and (b) $(h, h, 6+q)$ at 5.5 K with $q=0.796$. The magnetic field is applied along $[\bar{1} 1 0]$. 
(c) Magnetic field dependence of the peak-top intensities of the Bragg peaks at $(-\delta', 0, 6+q)$ and $(-\delta, -\delta, 6+q)$ at 5.5 K in the field increasing and decreasing processes, respectively. The inset in (c) schematically shows the peak positions. 
(d)  $\Delta \theta_{\text{PR}}$ dependence of the peak-top intensity at $(\delta, \delta, 6+q)$ at 2 T after domain selection. 
}
\label{fig:H110domain}
\end{figure}

By rotating the sample about the $c$ axis by $45^{\circ}$, we performed measurements for $H \parallel [\bar{1} 1 0]$. 
Figure \ref{fig:H110domain}(a) and \ref{fig:H110domain}(b) show the results of the reciprocal space scans along $(h, 0, 6+q)$ and $(h, h, 6+q)$ at several magnetic fields along $H \parallel [\bar{1} 1 0]$ at 5.5 K, where the phase II is realized at zero field with $q=0.796$. The detailed $T$ dependence of $q$ below $T_{\text{N}}^{\;\prime}$, which was not reported in the previous paper, is summarized in the SM~\cite{SM}. 
Figure \ref{fig:H110domain}(c) shows the magnetic field dependence of the intensities of the Bragg peaks at $(-\delta', 0, 6+q)$ and $(-\delta, -\delta, 6+q)$ at 5.5 K. 
The peak intensity of phase II starts to decrease above $\sim 0.7$ T, while that of phase III increases. This shows that the magnetic phase transition from phase II to phase III takes place by applying a magnetic field along $[\bar{1} 1 0]$. 
Since the Bragg peaks do not appear in the $(-h, h, 6+q)$ scan, these results again show that the cycloidal plane become perpendicular to the applied field.

When the cycloidal plane lies in the horizontal scattering plane, the $\Delta \theta_{\text{PR}}$ dependence of the intensity becomes flat as shown in Fig.~\ref{fig:H110domain}(d). 
This is because $F_{\pi\pi'}=0$ and $|F_{\pi\sigma'}|=|F_{\sigma\pi'}|$ are both satisfied. 
The former condition means that $\bm{m}_{\bm{q}}$ is parallel to the scattering plane. 
The latter condition, which is specific to the cycloid, is realized when the two orthogonal components of $\bm{m}_{\bm{q}}$ in the scattering plane have a phase difference of $\pi/2$. 
Otherwise, $|F_{\pi\sigma'}|$ is generally not equal to $|F_{\sigma\pi'}|$, and some structure proportional to $P_3$ should arise in the $\Delta \theta_{\text{PR}}$ dependence~\cite{SM}.

Normal domain selection occurs for  $H \parallel [0 1 0]$ and $H \parallel [\bar{1} 1 0]$ in phase II and III, respectively. 
When the field is applied along $[0 1 0]$ at 5.5 K in phase II, the intensities of the Bragg peak at $(0, \pm\delta', 6+q)$ vanish above $\sim 0.55$ T, whereas those at $(\pm\delta', 0, 6+q)$ increase. 
In the same manner, when the field is applied along $[\bar{1} 1 0]$ at 2.0 K in phase III, the intensities of the Bragg peak at $(\pm\delta, \mp\delta, 6+q)$ vanish above $\sim 0.35$ T, whereas those at $(\pm\delta, \pm\delta, 6+q)$ increase. 
These data are summarized in the SM~\cite{SM}.  
In these cases, the cycloidal domains are simply selected so that the cycloidal planes are perpendicular to the applied field. 
The uniform magnetization is induced along the field direction and the transverse conical structure is realized.


To compare the results obtained from RXD with the bulk property, we measured the $T$ dependence of magnetization at several magnetic fields applied along [110], which has not been reported previously. 
Figure \ref{fig:PhaseDiagram}(a) shows the $T$ dependence of magnetization in the form of $M/H$ for $H\parallel [110]$ measured at 0.1 T, 0.5 T, and 1.0 T. 
The results at 0.1 T and 0.5 T clearly exhibit an anomaly at the transition between phase II and III, where the cycloidal plane rotates by $45^{\circ}$. This anomaly is not detected in $M/H$ at 0.1 T for $H\parallel [100]$. 
At 1 T, which is above the critical field of phase II for $H\parallel [110]$, only the weak cusp anomaly at $T_{\text{N}}^{\;\prime}$ is detected. 
Below $T_{\text{N}}^{\;\prime}$, the sinusoidal order of phase I directly changes to the cycloidal order of phase III. 

The above results in this work are summarized in the phase diagrams in Figs.~\ref{fig:PhaseDiagram}(b) and \ref{fig:PhaseDiagram}(c) for $H\parallel [110]$ and $[100]$, respectively. 
The transition at $H^* \sim 2$ T for $H \parallel [1 0 0]$ reported in Refs.~\onlinecite{Maurya16} and \onlinecite{Kakihana17} is the transition from phase III to II, i.e., the rotation of the cycloidal plane, which is accompanied by the domain selection. 
The boundary below 1 T for $H \parallel [1 0 0]$ in phase II reported in Ref.~\onlinecite{Maurya16} can be attributed to the anomaly corresponding to the domain selection in phase II. 
The general response of the cycloidal order below  $T_{\text{N}}^{\;\prime}=7.5$ K is that the cycloidal plane prefers to be perpendicular to the external magnetic field. For $H \parallel [1 0 0]$ ($[1 1 0]$), phase II (III) is more stable in magnetic fields, and the cycloidal plane lies within the $[0 1 0]$--$[ 0 0 1]$ plane ($[\bar{1} 1 0]$--$[0 0 1]$ plane) with the uniform magnetization induced along the $[1 0 0]$ axis ($[1 1 0]$ axis).


When the cycloidal order is formed, only one mirror plane coinciding with the cycloidal plane remains. 
Another structure related by the four-fold rotation or by another mirror reflection, which is not preserved in the ordered state, gives the cycloidal structure of another domain. 
The mirror symmetry is preserved in the transverse conical state when a magnetic field is applied along the direction perpendicular to the mirror plane. 

When the transverse conical state is realized in magnetic fields, a lattice modulation with the same propagation vector is allowed by symmetry in such a way that the local polarization proportional to $\bm{e}_{ij} \times (\bm{S}_i \times \bm{S}_j)$ is induced, where $\bm{e}_{ij}$ represents the translation vector between $\bm{S}_i $ and $\bm{S}_j$ along the magnetic propagation vector $\bm{q}$~\cite{Katsura05,Arima07}. 
In the present case, a lattice modulation can be induced along the direction parallel to the uniform magnetization. 
This would give rise to a nonresonant scattering with the strong $\sigma$-$\sigma'$ component. 
However, no such signal has been detected as shown in Fig.~\ref{fig:H110domain}(d) and in the SM. 
Also in Fig.~\ref{fig:H010domain}(c), the intensities at $(\pm \delta', 0, 0)$ do not increase above 2 T, suggesting that no such additional scattering occurs. 
These results indicate that the spin-lattice coupling is tiny, possibly because of the $L=0$ orbital moment of Eu$^{2+}$ and the metallic nature of this compound. 
This suggests that the weak magnetic anisotropy could be due to the Fermi surface geometry, which is responsible for the magnetic exchange interaction and definitely reflects the tetragonal symmetry of the lattice\cite{Kakihana17}. 

From the phase diagram and the magnetization process, the transverse conical phase is expected to persist up to almost 15 T for $H \parallel ab$-plane. Nontrivial magnetic order such as the skyrmion lattice state is therefore unlikely to occur in this field direction.  
Magnetic structure study for $H \parallel [001]$ should be performed in future experiments.


In summary, we performed RXD using circularly polarized beam to clarify the helicity of the cycloidal magnetic structure in phases II and III of EuIrGe$_3$ at zero field. 
The magnetic structures, including the cycloidal helicity we observed for each magnetic domain, perfectly reflect the symmetry operations of the crystal structure with the point group $C_{4v}$. 
This result confirms that the cycloidal helicity is determined by the DM-type antisymmetric exchange interaction.

We also performed RXD to study the magnetic structure in magnetic fields applied along the [100] and [110] directions.  
When a magnetic field is applied along the [100] direction in Phase III, a phase transition occurs at $\sim2 $ T, at which the cycloidal plane rotates by $45^{\circ}$ to be perpendicular to the applied field. With the uniform component induced along the field direction, a transverse conical order is realized, which is the same structure as in phase II after the domain selection. 
In the same manner, when a magnetic field is applied along the [110] direction in Phase II, a phase transition occurs at $\sim 0.8$ T to the same transverse cycloid as in Phase III after the domain selection.

\acknowledgments
This work was supported by JSPS Grant-in-Aid for Scientific Research (B) (no. JP20H01854) and by JSPS Grant-in-Aid for Transformative Research Areas (Unveiling, Design, and Development of Asymmetric Quantum Matters, 23A202, no. JP23H04867). 
The synchrotron experiments were performed under the approval of the Photon Factory Program Advisory Committee (No. 2022G114). 
MT is supported by JST, the establishment of university fellowships towards the creation of science technology innovation, Grant No. JPMJFS2129.

\bibliography{EuIrGe3_1}


\clearpage
\begin{widetext}
\begin{center}
\textbf{\large{Supplemental Material}}
\end{center}
\vspace{5mm}

\begin{center}
\textbf{\large{Helicity Selection of the Cycloidal Order in Noncentrosymmetric EuIrGe$_3$}} \\
\vspace{2mm}
K. Kurauchi, T. Matsumura, M. Tsukagoshi, N. Higa, M. Kakihana, M. Hedo, T. Nakama, and Y. \={O}nuki
\end{center}
\vspace{20mm}
\end{widetext}

\renewcommand{\topfraction}{1.0}
\renewcommand{\bottomfraction}{1.0}
\renewcommand{\dbltopfraction}{1.0}
\renewcommand{\textfraction}{0.01}
\renewcommand{\floatpagefraction}{1.0}
\renewcommand{\dblfloatpagefraction}{1.0}
\setcounter{topnumber}{5}
\setcounter{bottomnumber}{5}
\setcounter{totalnumber}{10}

\renewcommand{\theequation}{S\arabic{equation}}
\renewcommand{\thefigure}{S\arabic{figure}}
\renewcommand{\thetable}{S-\Roman{table}}
\setcounter{section}{0}
\setcounter{equation}{0}
\setcounter{figure}{0}
\setcounter{page}{1}

\section*{Resonant X-ray Diffraction Experiment}
\begin{figure}[b]
\begin{center}
\includegraphics[width=8cm]{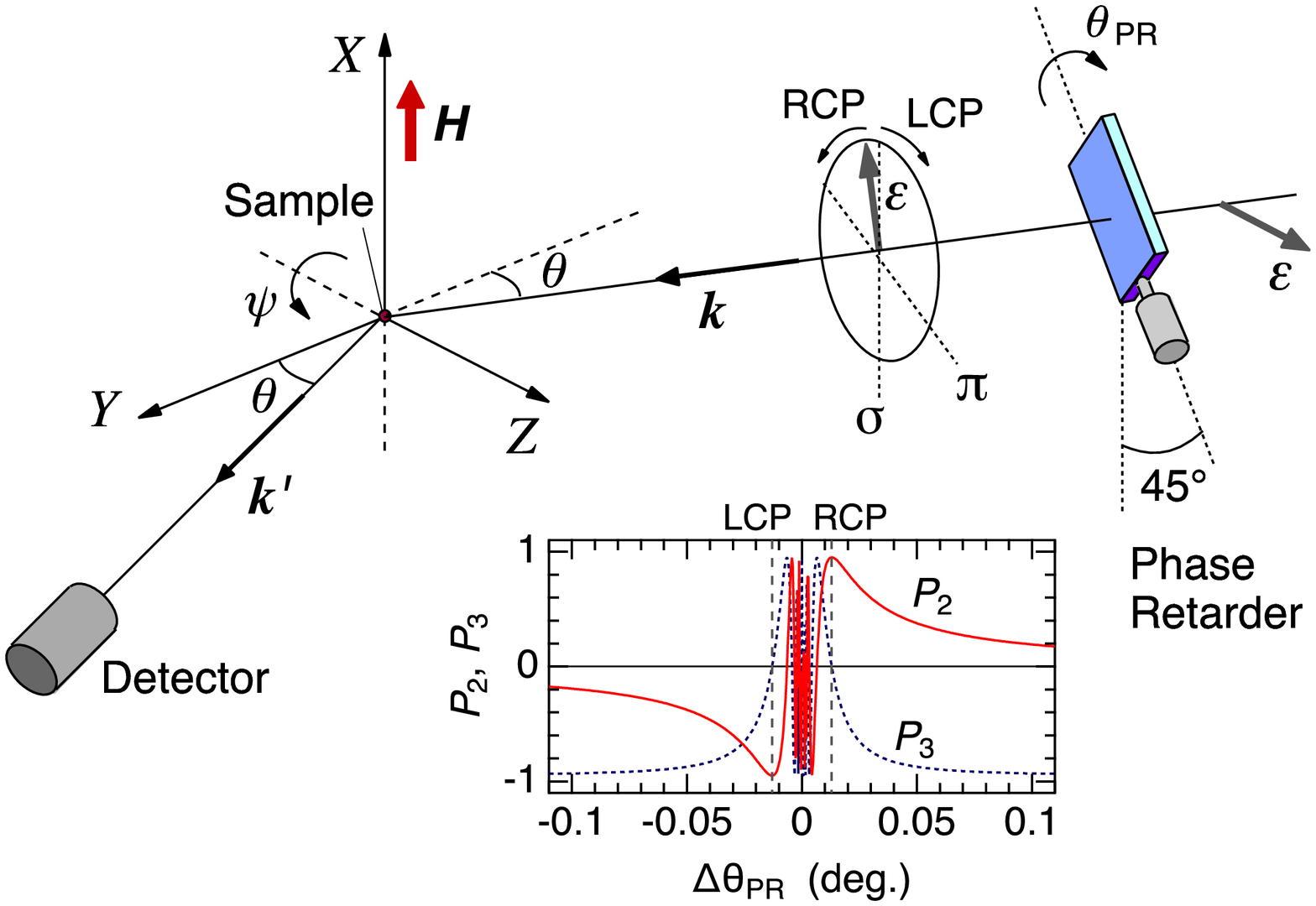}
\end{center}
\caption{
Schematic of the scattering geometry of the experiment with a phase retarder system inserted in the incident beam. 
Inset shows the $\Delta \theta_{\text{PR}}$ ($=\theta_{\text{PR}} - \theta_{\text{B}}$) dependence of the stokes parameters $P_2$ and $P_3$. 
}
\label{fig:ScattConfig}
\end{figure}

Figure~\ref{fig:ScattConfig} shows the scattering geometry of our RXD experiment. 
When the incident X-ray beam from the synchrotron source, which is polarized in the horizontal plane, passes through the diamond phase plate set near a Bragg angle, a phase difference occurs between the $\sigma$ and $\pi$ components with respect to the scattering plane tilted by $45^{\circ}$. 
The phase difference $\phi$ is proportional to $1/(\theta_{\text{PR}} - \theta_{\text{B}})$, where $\theta_{\text{B}}$ is the Bragg angle of the phase plate. 
It is therefore possible to tune the incident linear polarization to right-handed circular polarization (RCP) and left-handed circular polarization (LCP) by manipulating $\Delta\theta_{\text{PR}} = \theta_{\text{PR}} - \theta_{\text{B}}$ around the Bragg angle $\theta_{\text{B}}$. 
The polarization state of the X-ray after transmitting the phase plate is shown in Fig.~\ref{fig:ScattConfig} as a function of $\Delta\theta_{\text{PR}}$ using the Stokes parameters $P_2$ ($+1$ for RCP and $-1$ for LCP) and $P_3$ ($+1$ for $\sigma$ and $-1$ for $\pi$ linear polarization) [1]. 
In the horizontal scattering plane geometry in our experiment, $P_2= \sin \phi$ and $P_3 = -\cos \phi$. 
Near $\Delta\theta_{\text{PR}} = 0$ the beam becomes depolarized. 
$P_1$ ($+1$ for $45^{\circ}$ and $-1$ for $-45^{\circ}$ linear polarization) is zero in the present setup.

The scattering cross section can be expressed by using the four scattering amplitudes for $\sigma$-$\sigma'$, $\pi$-$\sigma'$, $\sigma$-$\pi'$, and $\pi$-$\pi'$ [1].  
\begin{align}
\Bigl( \frac{d\sigma}{d\Omega} \Bigr) &= 
\frac{1}{2} \bigl(\, |F_{\sigma\sigma'}|^2 + |F_{\sigma\pi'}|^2 + |F_{\pi\sigma'}|^2 + |F_{\pi\pi'}|^2 \,\bigr) \nonumber \\
 &\;\;\;\; + P_1 \text{Re} \bigl\{\, F_{\pi\sigma'}^*F_{\sigma\sigma'} + F_{\pi\pi'}^*F_{\sigma\pi'} \,\bigr\} \nonumber \\
 &\;\;\;\; + P_2 \text{Im} \bigl\{\, F_{\pi\sigma'}^*F_{\sigma\sigma'} + F_{\pi\pi'}^*F_{\sigma\pi'} \,\bigr\} 
 \label{eq:CrossSec1} \\
 &\;\;\;\; +  \frac{1}{2} P_3\bigl(\, |F_{\sigma\sigma'}|^2 + |F_{\sigma\pi'}|^2 - |F_{\pi\sigma'}|^2 - |F_{\pi\pi'}|^2 \,\bigr) 
 \,. \nonumber
\end{align}

\begin{figure}[b]
\begin{center}
\includegraphics[width=7.6cm]{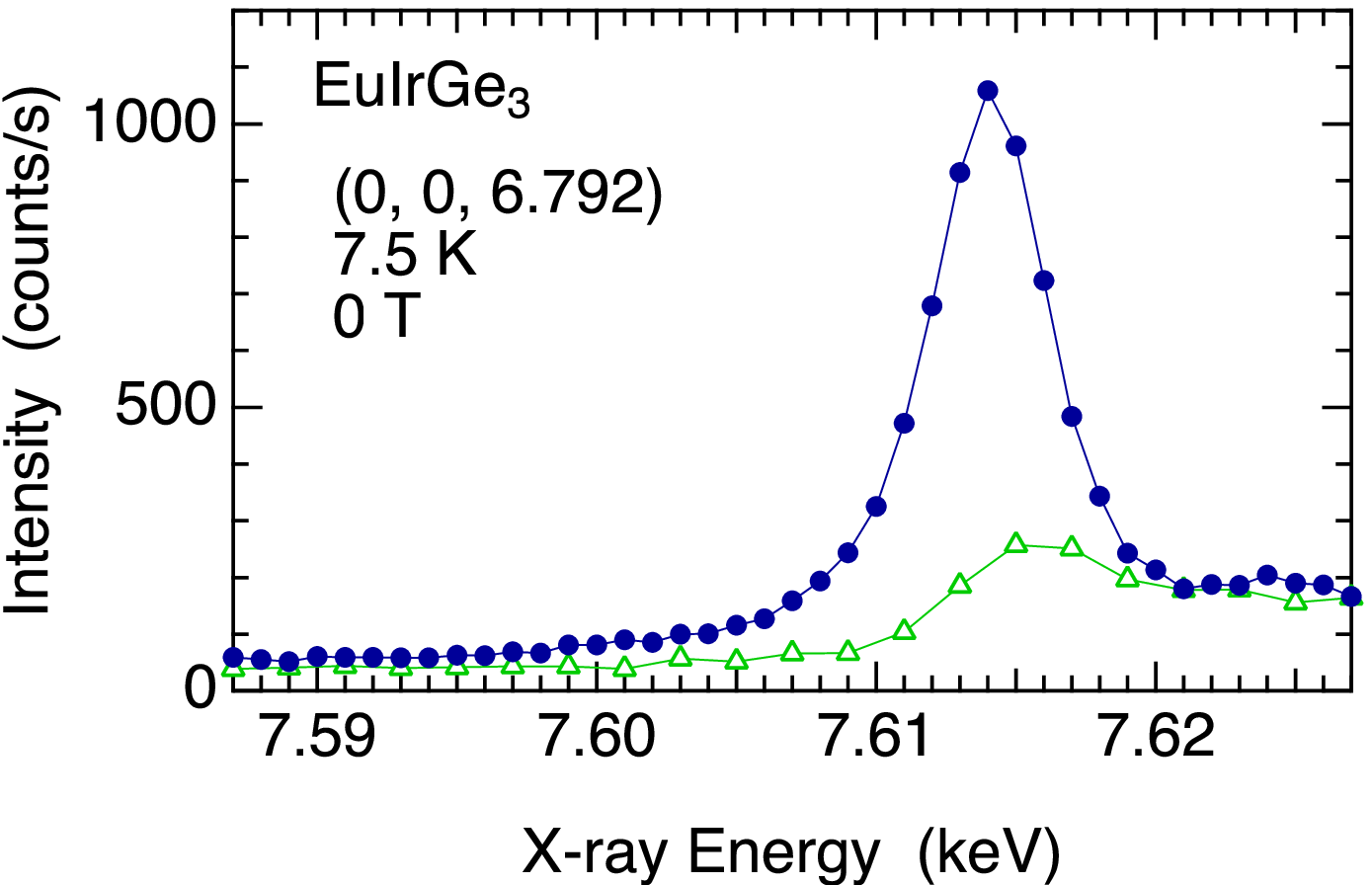}
\end{center}
\caption{
X-ray energy dependence of the $\bm{Q}=(0, 0, 6.792)$ resonance Bragg peak at 7.5 K in phase I. 
The triangles represent the background. 
}
\label{fig:Escan}
\end{figure}

The scattering amplitude $F_{\varepsilon\varepsilon'}(\omega)$ for the $E1$ resonant diffraction from a magnetic ordered structure is calculated by 
\begin{align}
F_{\varepsilon\varepsilon'}(\omega) &=  \alpha(\omega)  ( \bm{\varepsilon}' \times \bm{\varepsilon}) \cdot   \bm{Z}_{\text{M}} \;, \label{eq:S2}\\ 
\bm{Z}_{\text{M}} &= \sum_j \bm{m}_j \exp (-i \bm{Q}\cdot \bm{r}_j) \;, \label{eq:S3}
\end{align}
where $\bm{Z}_{\text{M}}$ represents the magnetic dipole structure factor for the scattering vector $\bm{Q}=\bm{k}' - \bm{k}$,  
$\bm{m}_j$ the magnetic dipole moment located at $\bm{r}_j$, and $\alpha(\omega)$ the spectral function of the $E1$ resonance [2,3].  
The energy spectrum of the resonant peak is shown in Fig.~\ref{fig:Escan}.

By assuming a model magnetic structure, we can compare the calculated cross section with the experimental $\Delta\theta_{\text{PR}}$ dependence of the intensity, which can often be a powerful tool to determine the magnetic structure and the helicity of the spiral orderings. 
With respect to the scattering from magnetic dipole orderings, $F_{\sigma\sigma'}$ vanishes. 
The asymmetric $\Delta\theta_{\text{PR}}$ dependence due to the $P_2$ term arises from the interference between $F_{\pi\pi'}$ and $F_{\sigma\pi'}$, which changes its sign when the helicity of the spiral is reversed.

\newpage
\section*{Temperature dependence of $\bm{q}$ at zero field}
\begin{figure}[b]
\begin{center}
\includegraphics[width=8cm]{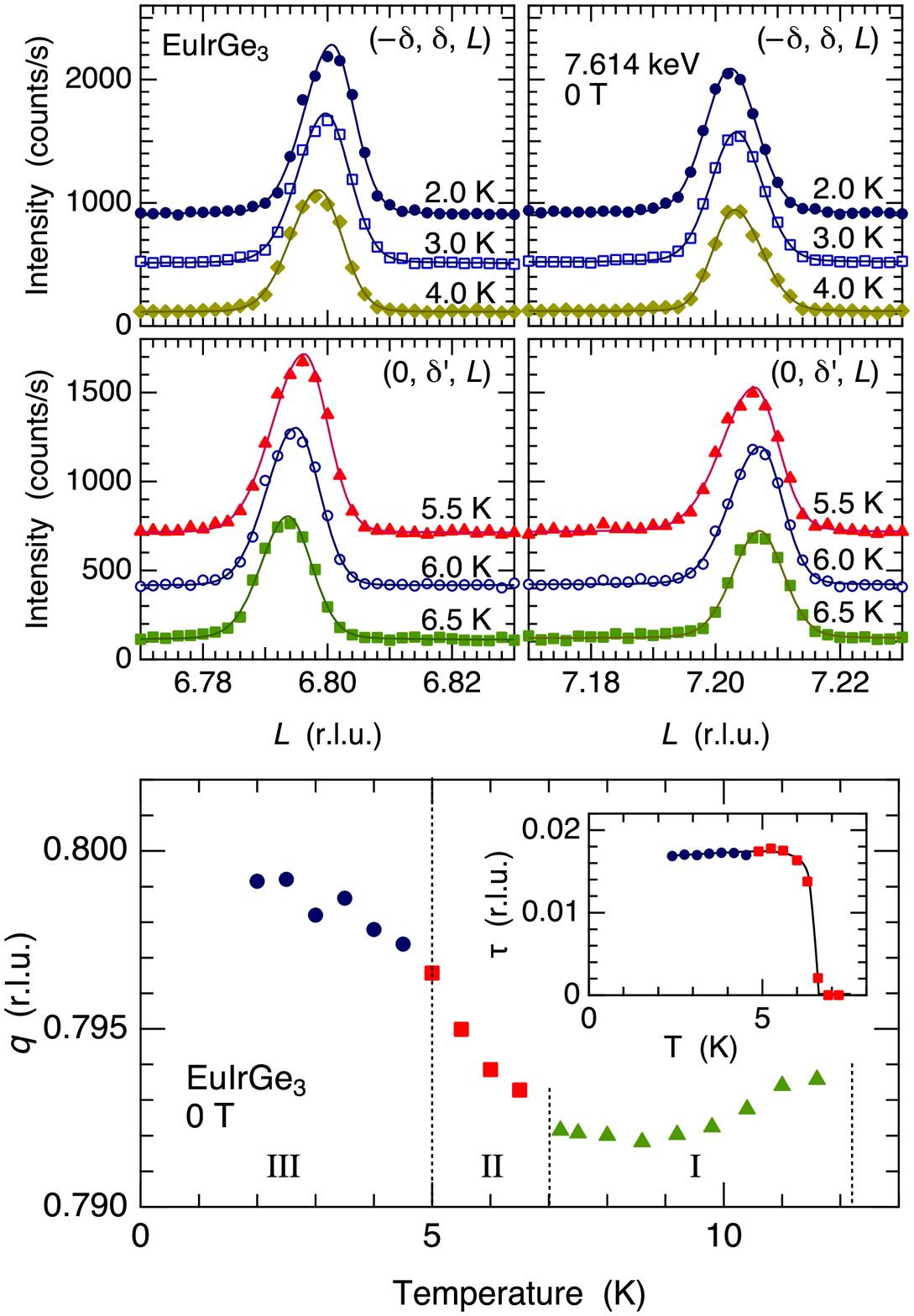}
\end{center}
\caption{
(top panels) Temperature dependence of the reciprocal scan profiles along the $(-\delta, \delta, L)$ and the $(0, \delta', L)$ lines in phase III and phase II, respectively. 
(bottom) Temperature dependence of the $q$ value of the peak position at $(0, 0, q)$, $(\delta', 0, q)$, and $(\delta, \delta, q)$ in phase I, II, and III, respectively. The inset shows the temperature dependence of $\tau=\delta'$ in phase II and $\tau=\sqrt{2}\delta$ in phase III, reproduced from Ref.~\onlinecite{Matsumura22}.
}
\label{fig:TdepQ0T}
\end{figure}

The detailed $T$ dependence of the $q$ value of the peak position at $(0, 0, q)$ in phase I, $(\delta', 0, q)$ in phase II, and $(\delta, \delta, q)$ in phase III, is shown in Fig.~\ref{fig:TdepQ0T}. The cycloidal pitch along the $c$ axis is given by $2\pi/q$. 
In our previous paper [4], only the $T$ dependence in phase I was reported. 
The $T$ dependence of the peak profiles along the $L$ direction around $L=6.8$ and 7.2 in reciprocal space, as shown in the upper panels, clearly shows that the $q$ value increases with decreasing temperature. It seems to approach 0.8 at the lowest temperature. 

\newpage
\section*{Normal domain selection in phase II and III}
\begin{figure}[b]
\begin{center}
\includegraphics[width=7.6cm]{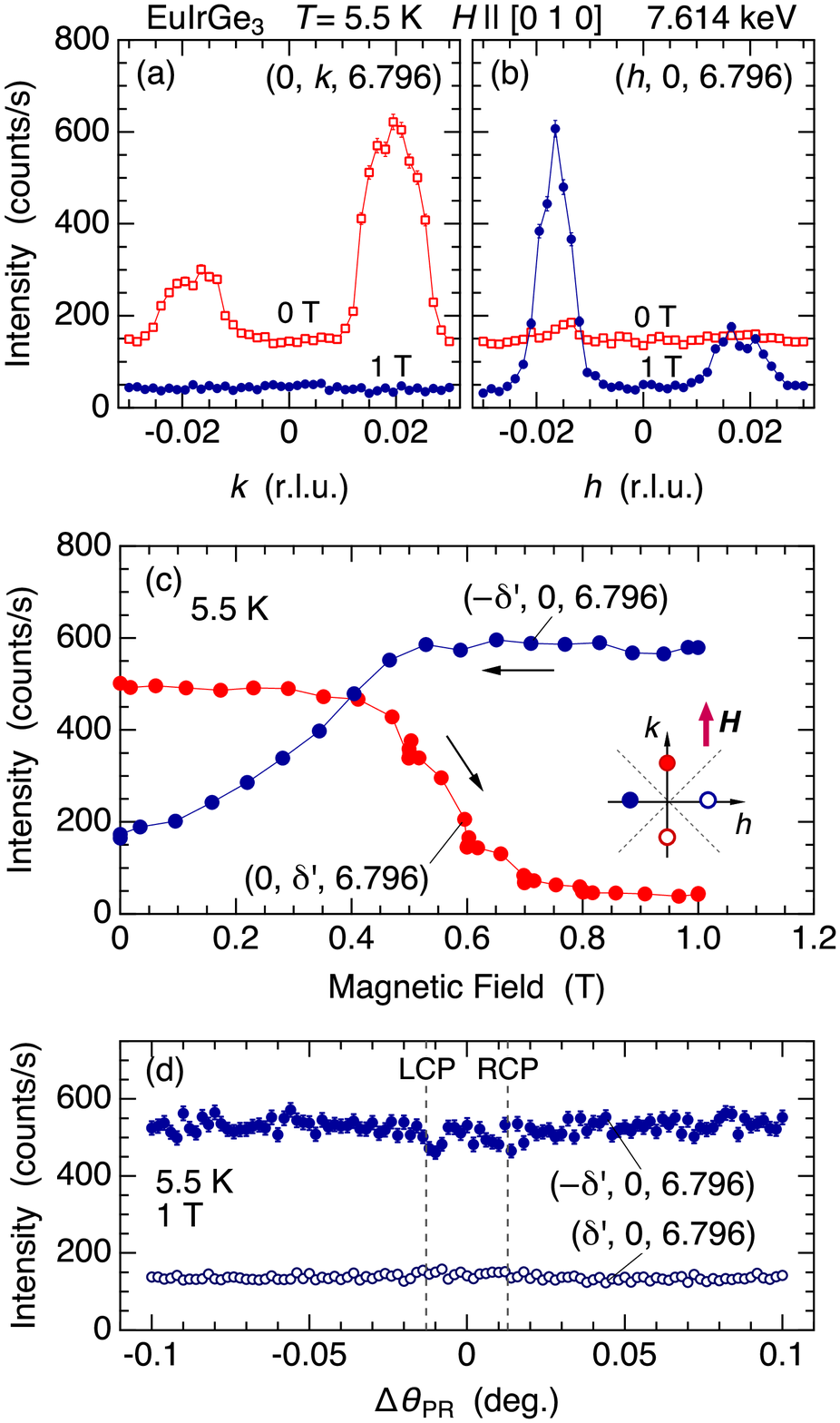}
\end{center}
\caption{(a) Peak profile of the $(0, k, 6+q)$ scan at 0 T and 1 T for $H \parallel [0 1 0]$ at 5.5 K in phase II. 
(b) Peak profile of the $(h, 0, 6+q)$ scan in the same condition as (a). 
(c) Magnetic field dependence of the peak-top intensity at $(0, \delta', 6+q)$ and $(-\delta', 0, 6+q)$ in the field increasing and decreasing process, respectively. 
(d)  $\Delta \theta_{\text{PR}}$ dependence of the peak-top intensity at $(\pm\delta', 0, 6+q)$ at 1 T after domain selection. 
}
\label{fig:Hdep010_5K}
\end{figure}

\begin{figure}[t]
\begin{center}
\includegraphics[width=7.6cm]{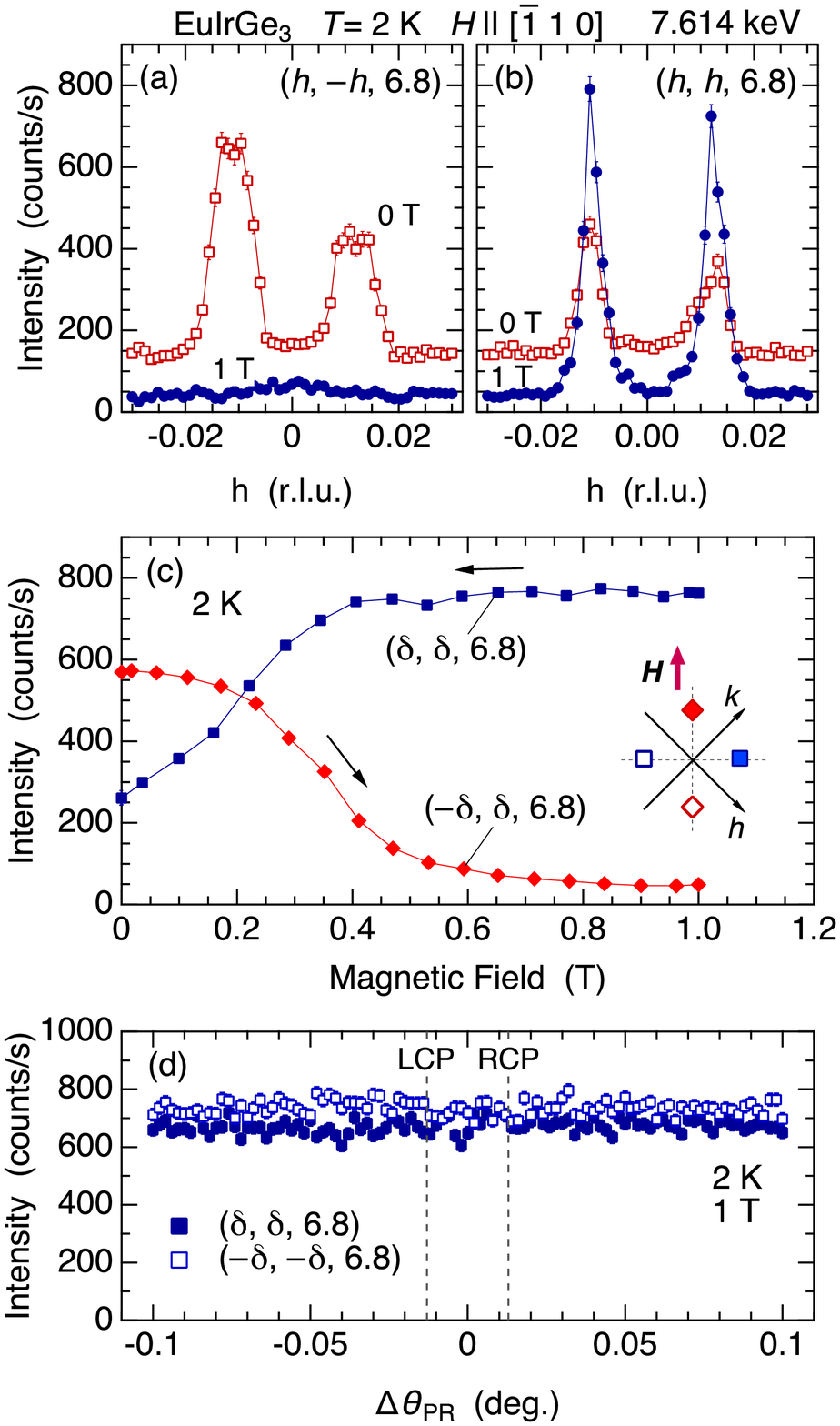}
\end{center}
\caption{(a) Peak profile of the $(h, -h, 6.8)$ scan at 0 T and 1 T for $H \parallel [\bar{1} 1 0]$ at 2 K in phase III. 
(b) Peak profile of the $(h, h, 6.8)$ scan in the same condition as (a). 
(c) Magnetic field dependence of the peak-top intensity at $(-\delta, \delta, 6.8)$ and $(\delta, \delta, 6.8)$ in the field increasing and decreasing process, respectively. 
(d)  $\Delta \theta_{\text{PR}}$ dependence of the peak-top intensity at $(\pm\delta, \pm\delta, 6.8)$ at 1 T after domain selection. 
}
\label{fig:Hdep110_2K}
\end{figure}

When the external magnetic field is applied along [010] in phase II at 5.5 K, or along [110] in phase III at 2 K, normal domain selection occurs without changing the order parameter. 
Figures \ref{fig:Hdep010_5K}(a) and \ref{fig:Hdep010_5K}(b) show the peak profiles of the $(0, k, 6.796)$ and $(h, 0, 6.796)$ scans, respectively, for $H \parallel [010]$ at 5.5 K in phase II. 
The $(0, \pm\delta', 6.796)$ peak existing at 0 T vanishes at 1 T and the $(\pm\delta', 0, 6.796)$ peak develops, resulting in the $\bm{q} \perp \bm{H}$ condition. 
The field dependences of the intensity of these peaks are shown in Fig.~\ref{fig:Hdep010_5K}(c). 
The boundary of the domain selection in the field increasing process is $\sim 0.55$ T, which is shown by the dotted line in Fig.~\ref{fig:PhaseDiagram}(c) in the main text. 
With respect to $H \parallel [\bar{1} 1 0]$ at 2 K in phase III, the results are shown in Figs.~\ref{fig:Hdep110_2K}(a), \ref{fig:Hdep110_2K}(b), and \ref{fig:Hdep110_2K}(c), and the boundary of the domain selection is shown by the dotted line in Fig.~\ref{fig:PhaseDiagram}(b) in the main text. 

The $\Delta \theta_{\text{PR}}$ dependences of the Bragg-peak intensity after the domain selection is shown in Fig.~\ref{fig:Hdep010_5K}(d) for the $(\pm\delta', 0, 6.796)$ peak and in Fig.~\ref{fig:Hdep110_2K}(d) for the $(\pm\delta, \pm\delta, 6.8)$ peak. 
The results are completely flat in both cases, indicating that the Fourier component $\bm{m}_{\bm{q}}$ is in the horizontal scattering plane and $F_{\pi\pi'}$ vanishes. There is no contribution from the $P_2$ term.  Since $|F_{\sigma\pi'}| = |F_{\pi\sigma'}|$, the $P_3$ term also vanishes. These results imply that the transverse conical state is realized with the cycloidal plane being perpendicular to the field and the uniform magnetization induced along the field direction. 
The flat data also show that no Thomson scattering from lattice modulation is induced, which is allowed by symmetry consideration as discussed in the main text. 
If the Thomson scattering exists, it would give rise to a $\sigma$-$\sigma'$ scattering and the increase in intensity at $\Delta \theta_{\text{PR}}$ half of the LCP and RCP positions.

\section*{Flat $\Delta \theta_{\text{PR}}$ dependence after domain selection}
The scattering-amplitude components in Eq.~(\ref{eq:S2}) are written as
\begin{align}
F_{\sigma\sigma'} &= 0 \;, \\
F_{\sigma\pi'} &= -m_y\cos\theta - m_z\sin\theta \;, \\
F_{\pi\sigma'} &= m_y\cos\theta - m_z\sin\theta \;, \\
F_{\pi\pi'} &= -m_x\sin 2\theta  \;, 
\end{align}
where $m_x$, $m_y$, and $m_z$ represent the $X$ ($\parallel \bm{k}\times\bm{k}'$), $Y$ ($\parallel \bm{k}+\bm{k}'$), and $Z$ ($\parallel \bm{k}'-\bm{k}$) component, respectively, of $\bm{m}_{\bm{q}}$ ($=\bm{Z}_{\text{M}}$) in Fig.~\ref{fig:ScattConfig}.
The spectral function $\alpha(\omega)$ was omitted. 
Since $F_{\sigma\sigma'}=0$, the $P_2$ term in Eq.~(\ref{eq:CrossSec1}) is proportional only to $\text{Im} \{ F_{\pi\pi'}^*F_{\sigma\pi'} \}$. 
After the cycloidal domain is selected by the external magnetic field, the magnetic moments lie in the $YZ$-scattering plane. 
Consequently, the $P_2$ term vanishes because $F_{\pi\pi'}=0$ ($m_x=0$). 
With respect to the $P_3$ term, it is usually expected to remain because $F_{\sigma\pi'}$ is generally not equal to $F_{\pi\sigma'}$ unless the scattering vector ($\parallel Z$-axis) is parallel to $\bm{m}_{\bm{q}}$, which is realized in phase I (see the data at 7.5 K in Fig.~\ref{fig:PRscans} in the main text). 
In the present case for the cycloidal order lying in the scattering plane after the domain selection, the two orthogonal components of $m_y$ and $m_z$ have a phase difference of $\pi/2$. Therefore, a special condition of $|F_{\sigma\pi'}|=|F_{\pi\sigma'}|$ is realized even when $\bm{m}_{\bm{q}}$ is not parallel to the scattering vector. Consequently, the $P_3$ term also vanishes. 

\vspace{5mm}
\small
\noindent
\newpage
\begin{minipage}[b]{0.47\textwidth}
\textbf{references} \\
\begin{enumerate}
\renewcommand{\labelenumi}{[\arabic{enumi}]}
\item S. W. Lovesey and S. P. Collins, \textit{X-ray Scattering and Absorption by Magnetic Materials} (Oxford, 1996). 
\item T. Nagao and J. Igarashi, Phys. Rev. B \textbf{72}, 174421 (2005). 
\item T. Nagao and J. I. Igarashi, Phys. Rev. B \textbf{74}, 104404 (2006). 
\item T. Matsumura, M. Tsukagoshi, Y. Ueda, N. Higa, A. Nakao, K. Kaneko, M. Kakihana, M. Hedo, T. Nakama, and Y. \={O}nuki, J. Phys. Soc. Jpn. \textbf{91}, 073703 (2022). 
\end{enumerate}
\end{minipage}
\end{document}